\documentclass[
prl,
superscriptaddress,
twocolumn,
amsmath,amssymb,
aps,
]{revtex4-2}

\usepackage{graphicx}
\usepackage{dcolumn}
\usepackage{bm}
\usepackage[caption=false]{subfig}

\usepackage[separate-uncertainty = true,multi-part-units=single]{siunitx}
\usepackage{color}

\newcommand{\OD}{\unit{OD}\,}

\begin{document}
%
%
%
\title{Driven shear flow in biological magneto-active fluids}%

\author{M.Marmol}
\affiliation{Aix Marseille Universit\'e, CEA, CNRS, BIAM, 13115 Saint Paul-Lez-Durance, France}
\affiliation{Claude Bernard Lyon 1 University, CNRS, Institut Lumi\`ere Mati\`ere, F-69622, Villeurbanne, France}
\author{C.Cottin-Bizonne}
\affiliation{Claude Bernard Lyon 1 University, CNRS, Institut Lumi\`ere Mati\`ere, F-69622, Villeurbanne, France}
\author{A.C\=ebers}
\affiliation{MMML Lab, Department of Physics, University of Latvia, Jelgavas-3, Riga, LV-1004, Latvia}
\author{D.Faivre}
\email{damien.faivre@lu.lv}
\affiliation{Aix Marseille Universit\'e, CEA, CNRS, BIAM, 13115 Saint Paul-Lez-Durance, France}
\affiliation{Department of Physics, University of Latvia, Jelgavas-3, Riga, LV-1004, Latvia}
\author{C.Ybert}
\email{Christophe.Ybert@univ-lyon1.fr}
\affiliation{Claude Bernard Lyon 1 University, CNRS, Institut Lumi\`ere Mati\`ere, F-69622, Villeurbanne, France}

%
\begin{abstract}
Active fluids made of powered suspended particles have unique abilities to self-generate flow and density structures. How such dynamics can be triggered and leveraged by external cues is a key question of both biological and applied relevance. Here we use magnetotactic bacteria to explore how chemotaxis and magnetotaxis -- leading, respectively, to positional and orientational responses -- combine to generate global scale flows. Such steady regime can be quantitatively captured by a magneto-active hydrodynamic model, while time-dependent magnetic driving unveils additional patterning complexity.
Overall, our findings shed light on how active fluids respond to the ubiquitous situation of multiple external information, also suggesting routes for their manipulation.

\end{abstract}
\maketitle

%
%
%
Suspensions of swimming microorganisms, or more broally self-propelled particles, give rise to so-called active fluids with fascinating properties \cite{Saintillan2018,Ramaswamy2019,Lauga2015}. These active fluids can self-generate flows, as in the case of bioconvection, where gravitaxis and swimming lead to density-driven 
convection patterns \cite{Bees2020, Qiu2022}, or in the so-called active turbulence, where the interplay of self-propulsion and interactions in dense suspensions gives rise to complex spatio-temporal flow patterns \cite{Alert2021}. Moreover, the interplay between swimming microbes and the surrounding flow field also results in heterogeneous spatial organization such as rheotactic drift in velocity gradients \cite{Marcos2012, Jing2020}. 

While active fluids owe their peculiar properties mostly to internal driving, other remarkable fluids are obtained with external driving, in particular for magnetic systems, which generate ferro- or magneto-rheological fluids \cite{Blums1997}. A fruitful route towards combining active and magnetic suspensions has therefore emerged, based on active spinners, magnetic rotors where magnetic particles in suspension are made active through rotating fields \cite{Jin2022, Junot2021}.
 Yet, a step further can be achieved with magnetotactic bacteria (MTB), which combine a swimming activity classical for bacterial systems with a magnetic response originating from intracellular ferrimagnetic nanoparticles of magnetite \cite{RevModPhys.96.021001}. Such systems already revealed novel hydrodynamic instabilities such as magneto-convection\,\cite{thery}, jetting \,\cite{waisbord}, or showed unique self-patterning mechanisms in the form of emerging stripes\,\cite{koessel} or Bose-Einstein like condensates\,\cite{bose}.

Here, we further explore the rich dynamics of this magnetic active biofluid. Classical biological swimmers indeed exhibit chemotaxis on top of their self-propulsion ability \cite{Raina2022}. Using MTB, we add the magnetic properties and demonstrate how chemical and magnetic responses combine to generate controllable global flow patterns and self-organizing phenomena reminiscent of large-scale canonical instabilities. The chemical - positional - cue breaks down the system homogeneity, with micro-organisms self-accumulating at a given location, while the magnetic cue breaks down the orientational isotropy. The combination of both phenomena gives rise to non-vanishing active stress that drives steady flows in predictable quantitative agreement with experiments.

%
%
%
\begin{figure}[h]
    \centering
    \includegraphics[width=0.35\textwidth]{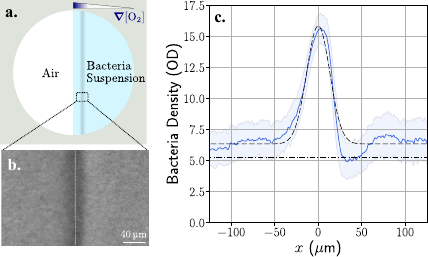}
    \caption{Experimental setup and Aerotactic band. 
    	(a) A \qty{95}{\micro\meter} thick and \qty{16}{mm} diameter cell is made with an adhesive spacer sandwiched between glass slides, half-filled with MTB suspension. 
    	(b) Aerotactic band formed once an oxygen gradient is established between the air reservoir and the suspension.
    	(c) Bacterial density profile across the band: vertically-averaged mean value (blue solid line); root-mean-squared error bars (blue shaded area); Gaussian fit $n(x) = n_\infty(1 + \Delta n \exp[-x^2/(2d^2)])$ with $n_\infty=\OD 6.3\simeq n_0$, $\Delta n = 1.5$, $d=\qty{13.3}{\micro\meter}$ (dot-dashed line), seeding density  $n_0 = \OD 5.3$ (long dashed line). 
    }
    \label{fig:Fig1}
\end{figure}

We used \textit{Magnetospirillum gryphiswaldense} (MSR-1) as bacterial species. These MTB share the dual ability to (i) orient along magnetic field lines; (ii) respond to an oxygen trigger for living purposes \cite{biophys2014,Popp2014,frankel97}. 
Fig.~\ref{fig:Fig1} shows the experimental setup
(see legend for details). 
Before sealing, the chamber is half filled with a suspension at a controlled MTB concentration $n_0$ (in the range of $\OD\numrange{1}{15}$, with $\OD 1 \equiv \qty{0.844E9}{bact.\per\milli\L}$). In these conditions, MTB consume dissolved oxygen and a gradient establishes with the air reservoir leading to the formation of a canonical aerotactic band \cite{biophys2014,bennet2014,codutti2019} as shown in Fig.~\ref{fig:Fig1}. Experiments made with calibrated image intensity allowed for determining the density profile $n(x)$ associated with such bands (see Supp. Mat for details). Overall, we describe the band by an excess density of Gaussian shape atop the background seeding level, with a typical band width $d\simeq\qty{15}{\micro\meter}$ and $\Delta n = (n_\mathrm{max.} - n_0)/n_0 \simeq 1.5$.

%
\begin{figure}
	\centering
	\includegraphics[width=0.48\textwidth]{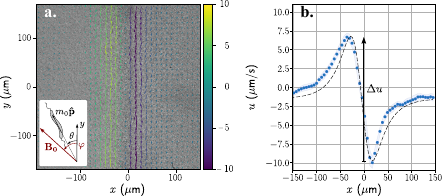}
	\caption{Magnetic field-induced shear flow. 
		(a) Aerotactic band image superimposed with PIV velocity field of bacteria showing the shear motion established around the density band ($n_0 = \OD 2$, $B_0 = \qty{5.6}{mT}$, $\varphi = \pi/4$). 
		Inset: Scheme for magnetic field and MSR-1 bacteria orientation.
		(b) $y$-averaged bacteria velocity profile; dashed line: theoretical prediction \eqref{eq:theo_high_field} with \qty{-1.5}{\micro\meter\per\second} offset to account for experimental background drift.
	}
	\label{fig:Fig2}
\end{figure}

\begin{figure}
	\centering
	\includegraphics[width=0.48\textwidth]{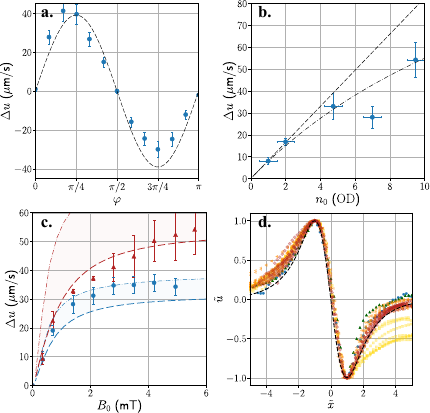}
	\caption{Shear flow characterization. 
		Evolution of the flow amplitude $\Delta u$: 
		(a) Magnetic field orientation dependency $\Delta u (\varphi)$ ($n_0 = \OD 4.7$, $B_0 = \qty{4}{mT}$); dashed line, same as in Fig.~\ref{fig:Fig2}b;
		(b) Density dependency  $\Delta u (n_0)$ ($B_0 = \qty{5.6}{mT} $, $\varphi=\pi/4$); dashed line, same as (a), dash-dotted line noise-free numerical solution (See Supp. Mat).
	(c) Magnetic field amplitude dependency, $\Delta u (B_0)$ $(\varphi=\pi/4$,  $n_0 = \OD 4.7$ (circle) or $n_0 = \OD 9.4$ (triangles)). Theoretical prediction including rotational noise with $T_\mathrm{eff.} = 4T_0$: Dash-dotted lines are dilute limit analytical calculation (see Supp. Mat for details) while the dashed lines frontiers correspond to full numerical solution. 
        (d) Normalized shear flow profiles: $\tilde{u} = 2(u-u_\mathrm{drift})/\Delta u$, $\tilde{x} = x/x_\mathrm{peak}$ with $2u_\mathrm{drift} = u_\mathrm{max} + u_\mathrm{min}$ and $2x_\mathrm{peak} = x(u_\mathrm{max}) - x(u_\mathrm{min})$. Data span $B_0 = \qtyrange{1.4}{5.6}{mT}$, $\varphi$ from $\pi/6$ to $\pi/3$ and  density in increasing order from blue to red: $\OD 1$ (circle), $\OD 2$ (upright triangle), $\OD 4.7$ (square), $\OD 7$ (left triangle), $\OD 9.5$ (diamond). Black dashed line is analytical prediction Eq. \eqref{eq:theo_high_field} with $d=\qty{15}{\micro\meter}$.
        }
	\label{fig:Fig3}
\end{figure}

Now, we submit the system to a uniform magnetic field of strength $B_0$ and orientation $\varphi$ with respect to the band axis (Fig.~\ref{fig:Fig2} a). Remarkably, as soon as $\varphi$ departs from the band axis $y$, a global shear motion develops around the band: the bacteria on both sides move parallel to the band and in opposite directions \footnote{See Movies SM1 and SM2 in Supp. Mat.}. 
This collective dynamics is quantified using particle image velocimetry (PIV) (see Supp. Mat. for details) and the measured $y$-averaged shear dynamics for $\varphi=\pi/4$ is shown in Fig.~\ref{fig:Fig2} b. Two opposite streams of bacteria extending \qtyrange{50}{100}{\micro\meter} from the band center are evidenced, separated by a shear zone whose total velocity amplitude is noted $\Delta u = \mathrm{max}[u({\scriptstyle x<0}) - u({\scriptstyle x>0})]$.
This shear motion can be tuned by varying the magnetic field angle $\varphi$, (Fig.~\ref{fig:Fig3}a) and a flow cancellation is observed for $0$ and $\pi/2$ together with the expected symmetries that yield a $\pi$-periodic response. This is in agreement with the fact that no such motion was ever reported in previous studies on magnetically assisted aerotaxis, where the magnetic field, typically chosen perpendicular to the band, serves as a proxy for the oxygen gradient direction.

 Fig.~\ref{fig:Fig3} gathers the quantitative characterization of the global dynamics. Setting the orientation $\varphi = \pi/4$ for which the velocity is the highest, a variation in bacterial density or field strength also controls the amplitude of the global motion that emerges. Denser suspensions show stronger responses despite a higher crowding, suggesting that the swimming activity powers the phenomenon (Fig.~\ref{fig:Fig3}b). Increasing the magnetic field also  enhances motion until saturation, qualitatively consistent with  bacterial orientation saturating upon full alignment with the  field (Fig.~\ref{fig:Fig3}c).
Beyond density, direction and amplitude, Fig.~\ref{fig:Fig3}d finally focuses on the shape of the velocity profile.
To compare all experimental conditions, the velocities are normalized by the amplitude $\Delta u/2$ after subtracting a possible residual background drift, estimated as $2u_\mathrm{drift} = u_\mathrm{max} + u_\mathrm{min}$. For position $x$, normalization is obtained from the peak position. Fig.~\ref{fig:Fig3}d gathers the resulting profiles for more than 30 configurations with different densities, field directions, and amplitudes, all collapsing onto a single 
 master curve, highlithing the robusteness of the  flow structure. Note that a few data do not fully relax to the quiescent state at large $\tilde{x}$, possibly due to a non-fully stationary bacterial band.

Before discussing the origins of this shear motion, let us stress that although derived from the bacterial suspension images, the velocity profiles reflect a large-scale fluid flow advecting the bacteria: experiments with added fluorescent colloids 
yielded identical velocity fields for swimming bacteria and for passive particles (See Supp. Mat.). 
%

%
%
%
Providing a full description of such flow generation by our magneto-active biological system should account for the joint bacteria probability function (position, orientation) \cite{koessel}, for the aerotactic response including motility reversal and oxygen transport \cite{codutti2019}, and for the coupling with hydrodynamic flows \cite{motorbrake}. We propose a simplified model, retaining minimal contributions and quantitatively describing the observed phenomena.
First, we observe that the aerotactic band does not show significant evolution under the various magnetic field forcings and their associated flow generation. Thus, we assume a fixed Gaussian shape bacterial number density $n(x)/ n_0 = 1 + \Delta n \exp(-x^{2}/[2d^{2}])$, of width $d$, reference level $n_0$ and amplitude $\Delta n$ (see Fig.~\ref{fig:Fig1}c). This simplifies the system in describing only the local bacterial orientation $\mathbf{\hat{p}}$ and the flow field $\mathbf{u}$.

A MSR-1 bacterium is modeled as a pusher swimmer with a negative force dipole strength $-\sigma_{0}$\,\cite{thery, Pierce2020} that carries a magnetic moment $\mathbf{m} = m_0 \mathbf{\hat{p}}$. Both swimming and magnetic dipoles share the same axis of direction. The rotational friction of the bacteria will be denoted $\xi_r$. In line with the measurement showing weak $z$-dependency, we restrict ourselves to a 2D $(x, y)$ system. Choosing for a length scale $L^*=d$, for velocities $U^* = n_0\sigma_0 d/\eta$ with $\eta$ the fluid viscosity, we have for time and stress the scales $T^*=\eta/(n_0\sigma_0)$ and $P^*=n_0\sigma_0$ respectively. Using $n_0$ to scale bacterial density, we obtain the following  dimensionless form of the flow equation:
\begin{equation}
(\Delta\mathbf{u}-\beta^{2}\mathbf{u})
- \boldsymbol{\nabla} p
+ \boldsymbol{\nabla}\cdot\boldsymbol{\overline{\Sigma}_p} = \mathbf{0};
\qquad \mathbf{\nabla}\cdot\mathbf{u}=0,
\label{eq:flow_full}
\end{equation}
where the second term accounts for the friction of the top and bottom walls through a Brinkmann approximation with $\beta^2 = 12d^2/h^2$ (see Supp. Mat. for the dimensional form and for the details of the derivation).
 
The particle stress tensor $\boldsymbol{\overline{\Sigma}_p}$ has passive, active, and magnetic contributions \cite{Alonso-Matilla2018, koessel}. For simplicity, we will restrict ourselves to a spherical bacterial shape. The particle stress term thus writes in dimensionless form:
\begin{equation}
	\boldsymbol{\overline{\Sigma}_p} = n 
	\left[
		\frac{5}{6}\eta_p \boldsymbol{\overline{\mathrm{E}}}
		- \left(\langle\mathbf{\hat{p}}\mathbf{\hat{p}}\rangle -\frac{\boldsymbol{\overline{\mathrm{I}}}}{3}\right)
		+ \alpha_m \left(\frac{\mathbf{\hat{b}}\langle\mathbf{\hat{p}}\rangle - \langle\mathbf{\hat{p}}\rangle\mathbf{\hat{b}}}{2}\right)
	\right]
\label{eq:stress_full}
\end{equation}
with $\boldsymbol{\overline{\mathrm{I}}}$ and $\boldsymbol{\overline{\mathrm{E}}}$ respectively the identity and strain-rate tensor $2\boldsymbol{\overline{\mathrm{E}}} = (\boldsymbol{\nabla}\mathbf{u} + [\boldsymbol{\nabla}\mathbf{u}]^T)$, $\mathbf{\hat{b}}$ the unit vector setting the magnetic field orientation, and with brackets denoting averaging over the local orientation distribution. 

The first term is also present for passive suspension, and $\eta_p =  n_0 \xi_r/\eta$ quantifies the local viscosity increase due to particles. The second term is the so-called active stress contribution associated with the pusher hydrodynamics nature of the swimming bacteria. The last term stands for the effect of magnetic torque whose magnitude is given by $\alpha_m= m_0B/\sigma_0$, which compares the magnetic and swimming contributions. 

This must be complemented by an equation for the bacterial orientation \cite{Alonso-Matilla2018, koessel} $\mathbf{\hat{p}}$:
\begin{equation}
	\mathbf{\dot{\hat{p}}} = 
	\left(
	\frac{1}{2}\boldsymbol{\nabla}\times\mathbf{u} + 
	\frac{\alpha_m}{\eta_p} \mathbf{\hat{p}}\times\mathbf{\hat{b}} + 
	\sqrt{\frac{\alpha_T}{\eta_p}}\boldsymbol{\Gamma(t)}
	\right)\times\mathbf{\hat{p}},
	\label{eq:orientation_full}
\end{equation}
where $\mathbf{\hat{p}}$ evolves due to flow vorticity, magnetic alignment, and rotational diffusion with $\boldsymbol{\Gamma(t)}$, a white noise verifying $\langle\Gamma_i(t)\Gamma_j(t')\rangle = 2\delta_{ij}\delta(t-t')$, and with $\alpha_T = k_BT/\sigma_0$, comparing thermal and swimming energies.

While the full equation system remains involved, much insights can be obtained in the strong magnetic field and low bacterial density limits for which $\alpha_m\gg \eta_p, \sqrt{\eta_p\alpha_T}$; $\eta_p\ll1$\footnote{Indeed, with typical values (see text) $\xi_r = \qty{7e-20}{\newton\meter\second}$, $\sigma_{0} = \qty{9e-19}{J}$ and $T = \numrange{1}{4}$ room temperature, we have for $\OD{1}$ and $B_0 = \qty{5.6}{mT}$: $\alpha_m \simeq 0.6$; $\eta_p\simeq\num{6e-2}$ and $\sqrt{\eta_p\alpha_T} = (\numrange{1.6}{3.2})\times10^{-2} $}.
Indeed, bacterial orientation becomes slaved to the external field one $\mathbf{\hat{p}} \simeq \mathbf{\hat{b}}$ and up to terms $o(\eta_p)$ associated with bacteria effect on viscosity, the flow becomes:
\begin{equation}
	\bigl[\partial_x^2 - \beta^2\bigr]u = -\frac{\sin(2\varphi)}{2}\partial_xn,
\end{equation}
where in agreement with experiments, $\mathbf{u}$ was taken as $u(x)\mathbf{\hat{y}}$, with $\mathbf{\hat{y}}$ the unit vector along the band direction. This equation has the following analytical solution:
\begin{equation}
	u(x) = -\sqrt{\frac{\pi}{32}}\Delta n \sin(2\varphi) F_\beta(x),
	\label{eq:theo_high_field}
\end{equation}
with the profile shape $F_\beta(x)$ being entirely controlled by the geometric factor $\beta$ comparing band width and slab thickness (see Supp. Mat for an explicit form). 

%
%

Remarkably, this extremely simplified model captures most of the experimental features of the shear flow (Fig.~\ref{fig:Fig3}a).
The effect of the magnetic field direction fully matches the $\sin2\varphi$ prediction while the flow velocity is expected to scale as $U^* = n_0\sigma_0 d/\eta$. 
As for bacteria density, the linear increase in $n_0$ well captures the general experimental trend  (Fig.~\ref{fig:Fig3}b) although our data may suggest some sub-linearity as we shall discuss shortly.

More quantitatively, setting $h=\qty{100}{\micro\meter}$ and the typical density band characteristics as measured $d=\qty{15}{\micro\meter}$, and $\Delta n = 1.5$, we obtain a quantitative agreement for a swimming force dipole $\sigma_{0} = \qty{9e-19}{J}$, in line with typical values for such microorganisms \cite{Drescher2011}, and with expectations based on the propelling velocity and the translational friction \cite{PICHEL2018, bennet2014}.
As we can see, the shape of the flow profile (Fig.~\ref{fig:Fig2}b) is captured by the analytical prediction, which closely matches the master curve evidenced for all experimental conditions (Fig.~\ref{fig:Fig3}d). Let us note that a small but systematic asymmetry seems present in the experimental data with a slightly longer flow extent on the left side of the profile. This coincides with the oxic region of the solution and may reflect a small anisotropy in the bacteria density, swimming properties \cite{bennet2014} or in the boundary provided by the free surface.

Even more remarkably, the model quantitatively matches the amplitudes of the magnetically-induced flow, both for the effect of $\varphi$ and of density $n_0$ (Fig.~\ref{fig:Fig3}a-b). For the latter, however, the simple linear prediction overestimates high density responses. Indeed, for $n_0 = \OD 10$, the dilute bacteria limit $\eta_p\ll1$ no longer holds as $\eta_p \simeq 0.6$ for $\xi_r=\qty{7e-20}{\N\m\s}$ \cite{PICHEL2018}.

To capture both the high density regime and the effect of finite magnetic field amplitude, flow and noise contributions to the orientation must be considered, requiring that Eqs. \eqref{eq:flow_full}--\eqref{eq:orientation_full} be solved numerically, taking for the magnetic moment $m_0 = \qty{1e-16}{A\per\meter\squared}$ \cite{bennet2014}, and using finite element method (see Supp. Mat. for details). 
First, we incorporate the finite-density effect while keeping the strong-field limit with negligible rotational noise: $\alpha_m\gg\sqrt{\eta_p\alpha_T}$. As shown in Fig.~\ref{fig:Fig3}b, this allows capturing quantitatively the sublinear evolution of the shear flow with $n_0$.
Finally, including the rotational noise effect, it is possible to fit the experimental response with respect to the field amplitude $B_0$ (Fig.~\ref{fig:Fig3}d), providing an effective temperature is accounted for, whose value is found of the order of $T_\mathrm{eff.} \simeq 4 T_0$. 
Indeed,  orientational noise likely includes active contributions, exceeding the bath temperature $T_0 = \qty{293}{K}$ and the value found here is consistent with previous estimates in other contexts \cite{Nadkarni.2013}.

Overall, our simple 1-dimensional hydrodynamic modeling  quantitatively captures the shear flow induced by a magnetic field acting on an aerotactic band of MTB. The combination of density heterogeneity, brought by the aerotactic response, and of alignment, induced by the external field, generates a controllable non-vanishing active stress driving  fluid flow throughout the system. Note that a key point of the  description is the decoupling between the band density and magnetic actuation. Indeed, North and South seekers  coexist in our sample and MSR-1 bacteria 
retain their pusher behavior upon motion reversal due to their bi-flagellated structure, leading to an orientational order in the suspension of nematic type. This may not hold true, in particular for species that are polar in their structure and/or in their magneto-aerotactic response, with their ability to form a band that depends upon field inversion with respect to the gradient direction \cite{biophys2014, RevModPhys.96.021001}. We expect this variety of magneto-aerotactic responses and of swimming characteristics will produce a rich set of collective effects.

%
%
%
%
%
%
%
%
%
\begin{figure}[!t]
    \centering
    \includegraphics[width=1\linewidth]{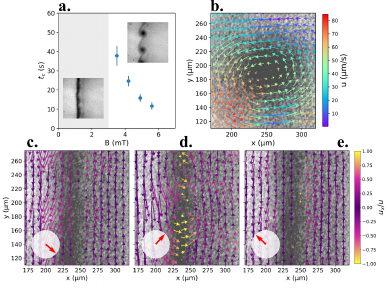}
    \caption{Rotating magnetic field inducing aerotactic band destabilization ($n_0=\OD 15$, $f=\qty{1}{Hz}$). 
    (a) Evolution of the characteristic time of destabilization $t_c$ with the magnetic field amplitude . 
    (b) Zoom on a bacteria patch at late stage, showing the associated vortex flow structure.
    (c-e) Early stage of aerotactic band breakdown: band images superimposed to PIV velocity field over the first period of field rotation ($B_0=\qty{5.6}{mT}$). The red arrow indicates the instantaneous orientation of the magnetic field.\label{fig:Fig4}}
\end{figure}
More complex dynamics can indeed be triggered by changing from a steady magnetic field to a rotating one, even for MSR-1 bacteria. 
Based on the above results, one may anticipate that for small frequency the shear flow magnitude and direction is simply modulated according to the instantaneous field orientation. 
Yet, we also expect the appearance of a bias due to the additional spinner characteristics of the bacteria provided by the rotating field. Indeed, a simple extension of our analytical prediction Eq.~\eqref{eq:theo_high_field} changes the $\sin2\varphi$-dependency into $(-\eta_p\omega + \sin2\omega t)$, where $\omega = \eta\Omega/(n_0\sigma_0)$ is the dimensionless form of the field angular velocity $\Omega$ (see Supp. Mat. for details).

After the initial phase following the above expectations, a more striking phenomenon develops, which now affects the band pattern. 
Past a characteristic time $t_c$ (see Supp. mat. for a quantitative definition), which depends on the exact field and density parameters, the straight band structure destabilizes and loses its $y$-invariance (Fig.~\ref{fig:Fig4}a).
Such destabilization can further evolve towards a full breakdown into a necklace of bacteria patches associated with flow vortices following the direction of the field rotation (Fig.~\ref{fig:Fig4}b and Movie SM3 in Supp. Mat.).

Note that destabilization of magnetic particles suspension into clusters under rotating field action was reported in different contexts \cite{Vach2017, Soni2019, Massana-Cid2021}. 
They however differ quite noticeably with the present phenomenon. 
Based on a competition of hydrodynamic and magnetic interactions, \cite{Massana-Cid2021} reported a destabilization of the whole suspension into clusters while here only the band seems to breakdown. 
More closely to our configuration, \cite{Soni2019} predicts that edge currents characteristic of spinner chiral active matter induce $y$-traveling excitations at the band edge which eventually yield to its breakdown. 
No such excitations were identified here, where we must keep in mind that the active swimming stress contribution considered earlier makes the dynamics significantly different from that of pure spinners, by modulating the band shearing.
Indeed, a series of snapshots at an early stage during the band evolution towards vortices (Fig.~\ref{fig:Fig4}c-e) shows how transverse displacements appear at specific phases of the rotation cycle, which will eventually yield the band breakdown and the vortex formation. 

%
%
Clearly, the description of the full dynamics of such mixed system will require further theoretical development, which may have to explicitly incorporate the microscopic response to the positional cue. Overall, this study demonstrates how the synergetic action of multiple cues can lead to original flow and pattern structures that can be shaped by the properties of the environment. Because this corresponds to a quite generic situation, we think this will open new perspectives on active fluids, both in biological or more physical contexts.


%
\begin{acknowledgments}
The authors thank F. Detcheverry, E. Gachon and E. Clément for fruitful discussions.
We acknowledge financial support from ANR-20-CE30-0034 BACMAG. DF also acknowledges financial support via the ERA Chair program of the European Union to the BioMagnetLink project (grant agreement
ID: 101187789). 
\end{acknowledgments}

\bibliography{Main}

\end{document}